\documentclass{IEEEtran} 
\IEEEoverridecommandlockouts

\usepackage{graphics}
\usepackage{epsfig}
\usepackage{times}
\usepackage{amsmath}
\usepackage{amssymb}
\usepackage{graphicx}
\usepackage{epstopdf}
\usepackage{epsfig}
\usepackage{enumerate}
\usepackage[noadjust]{cite}
\newtheorem{Theorem 1}{Theorem}
\newtheorem{Theorem 2}[Theorem 1]{Theorem}
\newtheorem{Theorem 3}[Theorem 1]{Theorem}
\newtheorem{Theorem 4}[Theorem 1]{Theorem}
\newtheorem{Theorem 5}[Theorem 1]{Theorem}
\newtheorem{Lemma 1}{Lemma}
\newtheorem{Lemma 2}[Lemma 1]{Lemma}
\newtheorem{Lemma 3}[Lemma 1]{Lemma}
\newtheorem{Definition 1}{Definition}
\newtheorem{Definition 2}[Definition 1]{Definition}
\newtheorem{Definition 3}[Definition 1]{Definition}
\newtheorem{Definition 4}[Definition 1]{Definition}
\newtheorem{Remark 1}{Remark}
\newtheorem{Remark 2}[Remark 1]{Remark}
\newtheorem{Remark 3}[Remark 1]{Remark}
\newtheorem{Corollary 1}{Corollary}
\newtheorem{Assumption 1}{Assumption}
\newtheorem{Assumption 2}[Assumption 1]{Assumption}
\usepackage{color}

\usepackage{stmaryrd}
\usepackage{stackengine}

\title{\LARGE \bf Dynamics Based Privacy Protection for Average Consensus on Directed Graphs}

\author{Huan~Gao,~\IEEEmembership{Student~Member,~IEEE,}
	and Yongqiang~Wang,~\IEEEmembership{Senior~Member,~IEEE}
	\thanks{Part of the results was accepted for presentation at 2018 IEEE Conference on Communications and Network Security (CNS) \cite{Huan2018CNS}. The work was supported in part by the National Science Foundation under Grant 1824014 and 1738902. }
	\thanks{Huan Gao and Yongqiang Wang are with the Department of Electrical and Computer Engineering, Clemson University, Clemson, SC 29634, USA {\tt\small \{hgao2, yongqiw\}@clemson.edu}}
}

\begin{document}

\maketitle

\begin{abstract}
Average consensus is key for distributed networks, with applications ranging from network synchronization, distributed information fusion, decentralized control, to load balancing for parallel processors. Existing average consensus algorithms require each node to exchange explicit state values with its neighbors, which results in the undesirable disclosure of sensitive state information. In this paper, we propose a novel average consensus approach for directed graphs which can protect the privacy of participating nodes' initial states without the assistance of any trusted third party or data aggregator. By leveraging the inherent robustness of consensus dynamics to embed privacy in random coupling weights between interacting nodes, our proposed approach can guarantee consensus to the exact value without any error. This is in distinct difference from differential-privacy based average consensus approaches which enable privacy through sacrificing accuracy in obtained consensus value. The proposed approach is able to preserve privacy even when multiple honest-but-curious nodes collude with each other. Furthermore, by encrypting exchanged information, the proposed approach can also provide privacy protection against inference by external eavesdroppers wiretapping communication links. Numerical simulations and hardware experiments on Raspberry Pi boards confirm that the algorithm is lightweight in computation and communication.
\end{abstract}

\section{Introduction}

Achieving average consensus is an important problem in distributed computing. For a distributed network of $N$ nodes interacting on a connected graph, average consensus can enable all nodes converge to the average of their initial values through iterations based on local interaction between neighboring nodes.

In recent years, average consensus has been extensively studied in distributed networks. Typical applications include load balancing (with divisible tasks) in parallel computing \cite{boillat1990load, cybenko1989dynamic}, network synchronization \cite{lynch1996distributed}, distributed information fusion \cite{scherber2004locally, xiao2005scheme}, and decentralized control \cite{olfati2004consensus, ren2005consensus}. To make all nodes converge to the average of their initial values, conventional average consensus algorithms require each node to exchange explicit state values with its neighbors. This leads to the disclosure of sensitive state information, which is undesirable in terms of privacy-preservation. In many collaborative applications such as smart grid, banking or health-care networks, privacy-preservation is crucial for encouraging participation in collaboration because individual nodes tend not to trade privacy for performance \cite{hoenkamp2011neglected}. For example, a group of individuals using average consensus to compute a common opinion may want to keep secret their individual personal opinions \cite{tsitsiklis1984problems}. Another example is power systems where multiple generators want to reach agreement on cost while keeping their individual generation information private \cite{zhang2011incremental}. Furthermore, exchanging information in the unencrypted plaintext form is vulnerable to attackers which try to steal information by hacking into communication links. With the increased number of reported attack events and the growing awareness of security, keeping data encrypted in communications has become the norm in many applications, particularly in many real-time sensing and control systems such as power systems and wireless sensor networks.

To enable privacy-preservation in average consensus, some results have been reported. One commonly used approach is differential privacy from the database literature \cite{huang2012differentially, nozari2015differentially, huang2015differentially, nozari2017differentially, katewa2017privacy}. However, differential-privacy based approaches do not guarantee the exact average value due to their fundamental trade-off between enabled privacy and computational accuracy \cite{wang2017differential, nozari2017differentially}. To guarantee computational accuracy, which is extremely important in sensor networks and cyber-physical systems, \cite{kefayati2007secure, manitara2013privacy, he2016private, mo2017privacy} proposed to inject additive correlated noise to exchanged messages, instead of uncorrelated noise used by differential-privacy. Observability based approaches have also been discussed to protect the privacy of multi-agent networks. The basic idea is to design the interaction topology so as to minimize the observability from a compromised agent, which amounts to minimizing its ability to infer the initial states of other network agents \cite{observability1, observability2}. However, observability based approaches cannot protect the privacy of the direct neighbors of the compromised agent. Moreover, both the correlated-noise based and the observability based approaches are vulnerable to external eavesdroppers who can steal exchanged information by hacking into the communication channels \cite{mo2017privacy}. Recently, the authors in \cite{ruan2017secure} proposed a new privacy-preserving approach to improving resilience to external eavesdroppers by encoding privacy in random coupling weights with the assistance of additive homomorphic encryption. However, all the aforementioned approaches require undirected or directed but balanced interactions and none of them are applicable to directed (potentially unbalanced) graphs due to the difficulties in constructing doubly stochastic weight matrices. It is worth noting that the approaches in \cite{dominguez2011distributed} and \cite{gharesifard2012distributed} for constructing doubly stochastic weight matrices require each node to explicitly send and reveal its weight and state to neighboring nodes, and hence are inapplicable when privacy preservation is required. 

In this paper, we address privacy-preserving average consensus under directed graphs that are not necessarily balanced. Building on the conventional push-sum based consensus algorithm, we enable privacy preservation by judiciously adding randomness in coupling weights and leverage the inherent robustness of push-sum to ensure consensus to the accurate value. The approach can also be easily extended to provide privacy protection against external eavesdroppers through incorporating partially homomorphic encryption. Both numerical simulations and hardware experiments are provided to verify the effectiveness and efficiency of the proposed approach.

\section{Preliminaries}

This section introduces some preliminaries of graph theory, the conventional push-sum algorithm, and the Paillier cryptosystem.


\subsection{Graph Theory}
Consider a network of $N$ nodes which is represented by a directed graph $\mathcal{G}=(\mathcal{V}, \, \mathcal{E})$ with node set $\mathcal{V}=\{1, 2, \ldots, N\}$. $\mathcal{E}\subset \mathcal{V} \times \mathcal{V}$ is the set of edges, whose elements are such that $(i, \, j) \in \mathcal{E}$ holds if and only if there exists a directed communication link from node $j$ to node $i$, i.e., node $j$ can send messages to node $i$. For notional convenience, we assume no self edges, i.e., $(i, \, i)\notin \mathcal{E}$ for all $i\in \mathcal{V}$. Each edge $(i, \, j)$ has an associated weight, $p_{ij}>0$. The out-neighbor set of node $i$, which represents the set of nodes that can receive messages from node $i$, is denoted as $\mathcal{N}_i^{out}=\{j \in \mathcal{V}  \, | \, (j, \, i)\in \mathcal{E}\}$. Similarly, the in-neighbor set of node $i$, which represents the set of nodes that can send messages to node $i$, is denoted as $\mathcal{N}_i^{in}=\{j \in \mathcal{V}  \, | \, (i, \, j)\in \mathcal{E}\}$. From the definitions, it can be obtained that $i \in \mathcal{N}_j^{out}$ and $j \in \mathcal{N}_i^{in}$ are equivalent to each other. Node $i$'s out-degree is denoted as $D_i^{out}=|\mathcal{N}_i^{out}|$ and its in-degree is denoted as $D_i^{in}=|\mathcal{N}_i^{in}|$, where $|\mathcal{S}|$ denotes the cardinality of set $\mathcal{S}$. In this paper, we focus on strongly connected graphs which are defined as follows:

\begin{Definition 1}
	A directed graph $\mathcal{G}$ is strongly connected if for any $i, j \in \mathcal{V}$ with $i\neq j$, there exists at least one directed path from $i$ to $j$ in $\mathcal{G}$, where the directed path respects the direction of the edges.
\end{Definition 1}

\subsection{The Conventional Push-Sum Algorithm}

Push-sum was introduced in \cite{kempe2003gossip} and \cite{benezit2010weighted} to achieve average consensus for nodes interacting on a directed graph that is not necessarily balanced (a balanced graph is a graph that satisfies $\sum_{j\in \mathcal{N}_i^{in}}p_{ij} = \sum_{j\in \mathcal{N}_i^{out}}p_{ji}$ for all $i$). In the conventional push-sum, $N$ nodes interact on a directed graph, with each node having an initial state $x_i^0$ ($i=1,2,\ldots,N$). Denote the average of all initial states as $\alpha={\sum_{j=1}^{N}x_j^0}/N$. The push-sum algorithm performs two iterative computations in parallel, and allows each node to asymptotically obtain the exact average of the initial values $\alpha$. The mechanism of the conventional push-sum algorithm is described as follows:

\noindent\rule{0.49\textwidth}{0.5pt}
\noindent\textbf{Algorithm 0:} Conventional push-sum algorithm

\vspace{-0.2cm}\noindent\rule{0.49\textwidth}{0.5pt}
\begin{enumerate}
	\item Each node $i$ initializes $s_i(0)=x_i^0$, $w_i(0)=1$, and $\pi_i(0)=s_i(0)/w_i(0)$. The coupling weight $p_{ij}$ associated with edge $(i, \, j)$ satisfies $p_{ij} \in (0,1)$ if $j\in \mathcal{N}_i^{in} \cup \{i\}$ holds and $p_{ij} =0$ otherwise. $p_{ij}$ also satisfies $\sum_{i=1}^N p_{ij}=1$ for $j=1, 2, \ldots, N$.
	\item At iteration $k$:
	\begin{enumerate}
		\item Node $i$ computes $p_{ji} s_i(k)$ and $p_{ji} w_i(k)$, and sends them to its out-neighbors $j \in \mathcal{N}_i^{out}$.
		\item After receiving $p_{ij} s_j(k)$ and $p_{ij} w_j(k)$ from its in-neighbors $j \in \mathcal{N}_i^{in}$, node $i$ updates $s_i$ and $w_i$ as follows:
		\begin{equation}\label{conventional_push_sum}
		\left\lbrace \begin{aligned}
		& \ s_i(k+1) = \sum_{j\in \mathcal{N}_i^{in} \cup \{i\}} p_{ij}s_j(k)\\
		& \ w_i(k+1) = \sum_{j\in \mathcal{N}_i^{in} \cup \{i\}} p_{ij}w_j(k)
		\end{aligned} \right.
		\end{equation}	
	    \item Node $i$ computes the estimate as $\pi_i(k+1)=s_i(k+1)/w_i(k+1)$.
	\end{enumerate}
\end{enumerate}
\vspace{-0.2cm}\rule{0.49\textwidth}{0.5pt}

For the sake of notational simplicity, the update rule in (\ref{conventional_push_sum}) can be rewritten in a more compact form as follows:
\begin{equation}\label{conventional_push_sum_vector_form}
\left\lbrace \begin{aligned}
& \ \mathbf{s}(k+1) =\mathbf{P} \mathbf{s}(k)\\
& \ \mathbf{w}(k+1) = \mathbf{P} \mathbf{w}(k)
\end{aligned} \right.
\end{equation}
where $\mathbf{s}(k)= [s_1(k), s_2(k), \ldots, s_N(k)] ^T$ and $\mathbf{w}(k)= [w_1(k), w_2(k), \ldots, w_N(k)]^T$, and $\mathbf{P}=[p_{ij}]$ with $p_{ij}$ defined in step 1) of Algorithm 0. According to Algorithm 0, we have $\mathbf{s}(0)= [x_1^0, x_2^0, \ldots, x_N^0] ^T$ and $\mathbf{w}(0)= \mathbf{1}$. It is also easy to obtain that matrix $\mathbf{P}$ is column stochastic since its column sums are equal to one, i.e., $\sum_{i=1}^N p_{ij}=1$ for $j=1, 2, \ldots, N$. 

At iteration $k$, each node calculates the ratio $\pi_i(k+1)=s_i(k+1)/w_i(k+1)$, which is used to estimate the average value $\alpha={\sum_{j=1}^{N}x_j^0}/N$. Since the directed graph $\mathcal{G}$ is assumed to be strongly connected, $\mathbf{P}^k$ converges to a rank-$1$ matrix exponentially fast \cite{Seneta_Markov_book, fill1991eigenvalue}. Defining $\mathbf{P}^\infty$ as the limit of $\mathbf{P}^k$ when $k \rightarrow \infty$, we have that $\mathbf{P}^\infty$ is of the form $\mathbf{P}^\infty=\mathbf{v} \mathbf{1}^T $ where $\mathbf{v}=[v_1, v_2, \ldots, v_N]^T$. Making use of the facts $\mathbf{s}(k)=\mathbf{P}^k \mathbf{s}(0)$ and $\mathbf{w}(k)=\mathbf{P}^k \mathbf{w}(0)$, we have
\begin{equation}\label{conventional_push_sum_convergence}
\begin{aligned}
\pi_i(\infty)= & \frac{s_i(\infty)}{w_i(\infty)}=\frac{ [\mathbf{P}^\infty \mathbf{s}(0)]_i}{[\mathbf{P}^\infty \mathbf{w}(0)]_i}= \frac{v_i \sum_{j=1}^{N} s_j(0)}{v_i \sum_{j=1}^{N} w_j(0)}\\ 
= & \frac{\sum_{j=1}^{N}x_j^0} {N}=\alpha
\end{aligned}
\end{equation}
where $[\mathbf{P}^\infty \mathbf{s}(0)]_i$ and $[\mathbf{P}^\infty \mathbf{w}(0)]_i$ represent the $i$-th element of vectors $\mathbf{P}^\infty \mathbf{s}(0)$ and $\mathbf{P}^\infty \mathbf{w}(0)$, respectively. Therefore, all estimates $\pi_1(k), \pi_2(k), \ldots, \pi_N(k)$ asymptotically converge to the average value $\alpha={\sum_{j=1}^{N}x_j^0}/N$.

\subsection{Paillier Cryptosystem}
Our method to protect privacy against external eavesdroppers wiretapping communication links is to encrypt exchanged messages. To this end, we combine Paillier cryptosystem \cite{paillier1999public} with consensus dynamics. The Paillier cryptosystem is a public-key cryptosystem which employs a pair of keys: a public key and a private key (also called secret key). The public key is distributed publicly and can be used by any person to encrypt a message, but such a message can only be decrypted by the private key. Since Paillier cryptosystem does not need the assistance of a trusted third party for key management, it is applicable in open and dynamic networks. 

The Paillier cryptosystem includes three algorithms as follows:

\noindent\rule{0.49\textwidth}{0.5pt}
\noindent\textbf{Paillier cryptosystem}

\vspace{-0.2cm}\noindent\rule{0.49\textwidth}{0.5pt}
\noindent\textbf{Key generation:}
\begin{enumerate}
	\item Choose two large prime numbers $p$ and $q$ of equal bit-length and compute $n=pq$.
	\item Let $g=n+1$.
	\item Let $\lambda=\phi(n)=(p-1)(q-1)$, where $\phi(\cdot)$ is the Euler's totient function.
	\item Let $\mu=\phi(n)^{-1}\mod n$ which is the modular multiplicative inverse of $\phi(n)$.
	\item The public key $k^p$ for encryption is $(n,g)$.
	\item The private key $k^s$ for decryption is $(\lambda,\mu)$.   
\end{enumerate}

\noindent\textbf{Encryption ($c={E}(m)$):}

Recall the definitions of $\mathbb{Z}_n=\{z \, | \, z\in\mathbb{Z},0\le z<n\}$ and $\mathbb{Z}_n^*=\{z \, | \,z\in\mathbb{Z},0\le z<n, \gcd(z,n)=1\}$ where
$\gcd(a, b)$ is the greatest common divisor of $a$ and $b$.
\begin{enumerate}
	\item Choose a random $r\in\mathbb{Z}_n^*$.
	\item The ciphertext is given by $c=g^m\cdot r^n\mod n^{2} $, where $m\in\mathbb{Z}_n, c\in\mathbb{Z}_{n^{2}}^*$.
\end{enumerate}

\noindent\textbf{Decryption ($m={D}(c)$):}
\begin{enumerate}
	\item Define the integer division function $L(\mu)=\frac{\mu-1}{n}$.
	\item The plaintext is $m=L(c^\lambda\mod n^{2})\cdot \mu \mod n $.
\end{enumerate}
\vspace{-0.2cm}\noindent\rule{0.49\textwidth}{0.5pt}

\section{The Privacy-Preserving Algorithm and Performance Analysis}

In this section, we first show that the conventional push-sum algorithm does not preserve privacy, and then we propose our privacy-preserving average consensus algorithm for directed graphs. To this end, we first introduce the attack model and the definition of privacy.

\begin{Definition 2}\label{honest_but_curious_definition}
	An honest-but-curious adversary is a node who follows all protocol steps correctly but is curious and collects received data in an attempt to learn the initial value of other participating nodes.
\end{Definition 2}

\begin{Definition 3}\label{privacy_preservation_definition}
	For a distributed network of $N$ nodes, the privacy of initial value $x_i^0$ of node $i$ is preserved if $x_i^0$ cannot be estimated by honest-but-curious adversaries with any accuracy.
\end{Definition 3}

Definition \ref{privacy_preservation_definition} requires that honest-but-curious adversaries cannot even find a range for a private value and thus is more stringent than the privacy definition considered in \cite{liu2006random, han2010privacy, cao2014privacy} which defines privacy preservation as the inability of an adversary to {\it uniquely} determine the protected value. It is worth noting that by a finite range, we mean lower and upper bounds that an adversary may obtain based on accessible information. We do not consider representation bounds caused by the finite number of bytes that can be used to represent a number in a computer.

We first show that the conventional push-sum algorithm is not privacy-preserving. According to (\ref{conventional_push_sum}) and (\ref{conventional_push_sum_vector_form}), an honest-but-curious node $i$ will receive two messages $p_{ij}s_j(0)$ and $p_{ij}w_j(0)$ from its in-neighbor node $j\in \mathcal{N}_i^{in}$ after the first iteration $k=0$. Then node $i$ can uniquely determine the initial value $x_j^0$ by $x_j^0 =s_j(0) = \frac{p_{ij}s_j(0)} {p_{ij}w_j(0)}$ using the fact $w_j(0)=1$. So an honest-but-curious node can infer the initial values of all its in-neighbors, meaning that the conventional push-sum algorithm cannot provide protection against honest-but-curious attackers.

\subsection{Privacy-Preserving Average Consensus Algorithm}

The above analysis shows that using the same coupling weight $p_{ij}$ for both $p_{ij}s_j(0)$ and $p_{ij}w_j(0)$ leads to the disclosure of the initial state value. Motivated by this observation, we introduce a novel privacy-preserving average consensus algorithm that uses different time-varying coupling weights for $s_i$ and $w_i$ for iterations $k=0, \ldots, K$, where $K$ can be any non-negative integer.

\noindent\rule{0.49\textwidth}{0.5pt}
\noindent\textbf{Algorithm 1:} Privacy-preserving average consensus algorithm

\vspace{-0.2cm}\noindent\rule{0.49\textwidth}{0.5pt}
\begin{enumerate}
	\item Positive integer $K$ and parameter $\varepsilon\in (0, \, 1)$ are known to every node. Each node $i$ initializes $s_i(0)=x_i^0$, $w_i(0)=1$, and $\pi_i(0)=s_i(0)/w_i(0)$.
    \item At iteration $k$:
          \begin{enumerate}
          	\item If $k \leq K$, node $i$ generates one set of random coupling weights $\big\{p_{ji}^s(k) \in \mathbb{R} \, \big| \, j\in \mathcal{N}_i^{out} \cup \{i\} \big\}$ with the sum of this set equal to $1$, and assigns $p_{ii}^w(k)=1$ and $ p_{ji}^w(k)=0$ for $j\in \mathcal{N}_i^{out}$; otherwise, node $i$ generates one set of random coupling weights $\big\{p_{ji}^s(k) \in  (\varepsilon, \, 1) \, \big| \, j\in \mathcal{N}_i^{out} \cup \{i\} \big\}$ satisfying the sum $1$ condition, and assigns $p_{ji}^w(k)=p_{ji}^s(k)$ for $j\in \mathcal{N}_i^{out} \cup \{i\}$. For $j\notin \mathcal{N}_i^{out} \cup \{i\}$, node $i$ always sets $p_{ji}^s(k)$ and $p_{ji}^w(k)$ to $0$.
          	\item Node $i$ computes $p_{ji}^s(k) s_i(k)$ and $p_{ji}^w(k) w_i(k)$ for $j\in \mathcal{N}_i^{out}$, and sends them to node $j$.
          	\item After receiving $p_{ij}^s(k) s_j(k)$ and $p_{ij}^w(k) w_j(k)$ from its in-neighbors $j \in \mathcal{N}_i^{in}$, node $i$ updates $s_i$ and $w_i$ in the following way
          	\begin{equation}\label{Algorithm_I_s_w_update}
          	\left\lbrace \begin{aligned}
          	& s_i(k+1) = \sum_{j\in \mathcal{N}_i^{in} \cup \{i\} } p_{ij}^s(k) s_j(k)\\
          	& w_i(k+1) = \sum_{j\in \mathcal{N}_i^{in} \cup \{i\} } p_{ij}^w(k) w_j(k)
          	\end{aligned} \right.
          	\end{equation}	
          	\item Node $i$ computes the estimate as $\pi_i(k+1)=s_i(k+1)/w_i(k+1)$.
          \end{enumerate}
\end{enumerate}
\vspace{-0.2cm}\rule{0.49\textwidth}{0.5pt}

Following Algorithm 1, the dynamics in (\ref{Algorithm_I_s_w_update}) can be summarized as follows:
\begin{equation}\label{our_algorithm_I}
\left\lbrace \begin{aligned}
& \ \mathbf{s}(k+1) =\mathbf{P}_{s}(k) \mathbf{s}(k)\\
& \ \mathbf{w}(k+1) = \mathbf{P}_{w}(k) \mathbf{w}(k)
\end{aligned} \right.
\end{equation}
where $\mathbf{s}(k)= [s_1(k), s_2(k), \ldots, s_N(k)] ^T$ and $\mathbf{w}(k)= [w_1(k), w_2(k), \ldots, w_N(k)]^T$, and the $ij$-th entries of $\mathbf{P}_{s}(k)$ and $\mathbf{P}_{w}(k)$ are the coupling weights $p_{ij}^s(k)$ and $p_{ij}^w(k)$, respectively. For the initial iteration, we have $\mathbf{s}(0)= [x_1^0, x_2^0, \ldots, x_N^0] ^T$ and $\mathbf{w}(0)= \mathbf{1}$. Under Algorithm 1, we have that $\mathbf{P}_{s}(k)$ and $\mathbf{P}_{w}(k)$ in (\ref{our_algorithm_I}) are time-varying, column stochastic, and satisfy $\mathbf{P}_{s}(k) \neq \mathbf{P}_{w}(k)$ for $k \leq K$ and $\mathbf{P}_{s}(k) = \mathbf{P}_{w}(k)$ for $k \geq K+1$. Note that $\mathbf{P}_{w}(k) = \mathbf{I}$ holds for $k=0,1,\ldots, K$. 

Define transition matrices as follows
\begin{equation}\label{transition_matrix}
\left\lbrace \begin{aligned}
& \ \mathbf{\Phi}_s(k:t) =\mathbf{P}_{s}(k) \cdots\mathbf{P}_{s}(t)\\
& \ \mathbf{\Phi}_w(k:t) =\mathbf{P}_{w}(k) \cdots\mathbf{P}_{w}(t)
\end{aligned} \right.
\end{equation}
for all $k$ and $t$ with $k \geq t$, where $\mathbf{\Phi}_s(k:k)=\mathbf{P}_{s}(k)$ and $\mathbf{\Phi}_w(k:k)=\mathbf{P}_{w}(k)$. Then the system dynamics in (\ref{our_algorithm_I}) can be rewritten as:
\begin{equation}\label{Algorithm_I_first_half}
\left\lbrace \begin{aligned}
& \mathbf{s}(K+1) = \mathbf{\Phi}_s(K:0) \mathbf{s}(0)\\
& \mathbf{w}(K+1) = \mathbf{\Phi}_w(K:0) \mathbf{w}(0)=\mathbf{1}\\
\end{aligned} \right.
\end{equation}
and
\begin{equation}\label{Algorithm_I_second_half}
\left\lbrace \begin{aligned}
& \mathbf{s}(k+1) = \mathbf{\Phi}_s(k:K+1) \mathbf{s}(K+1)\\
& \mathbf{w}(k+1) = \mathbf{\Phi}_w(k:K+1) \mathbf{w}(K+1)\\
\end{aligned} \right.
\end{equation}
	for $k \geq K+1$. In (\ref{Algorithm_I_first_half}) we used the facts $\mathbf{w}(0)=\mathbf{1}$ and $\mathbf{\Phi}_w(K:0)= \mathbf{I}$ (note that $\mathbf{P}_{w}(k) = \mathbf{I}$ for $k=0,1,\ldots, K$).

\subsection{Convergence Analysis}

Next we show that Algorithm 1 can guarantee convergence to the exact average value of all initial values. We will also analyze the convergence rate of Algorithm 1. Following the the convergence definition in \cite{nedic2017achieving} and \cite{nedich2016geometrically}, we define the convergence rate to be at least $\gamma \in (0, \, 1)$ if there exists a positive constant $C$ such that $\big\|\boldsymbol{\pi}(k) - \alpha \mathbf{1} \big\| \leq C \gamma ^k$ holds for all $k$, where $\boldsymbol{\pi}(k)=[\pi_1(k), \ldots, \pi_N(k)]^{T}$ and $\alpha={\sum_{j=1}^{N}x_j^0}/N$ is the average value. Note that a smaller $\gamma$ means a faster convergence. Before analyzing the convergence rate of Algorithm 1, we first introduce the following lemma.

\begin{Lemma 1}
	Consider a distributed network of $N$ nodes interacting on a strongly connected graph $\mathcal{G}=(\mathcal{V}, \, \mathcal{E})$. Under Algorithm 1, each node $i$ has $w_i(k)=1$ for $k \leq K+1$ and $w_i(k) \geq \varepsilon^N$ for $k \geq K+2$.
\end{Lemma 1}
{\it Proof}: The proof is given in the Appendix. \hfill{$\blacksquare$}

\begin{Theorem 1}
	Consider a distributed network of $N$ nodes interacting on a strongly connected graph $\mathcal{G}=(\mathcal{V}, \, \mathcal{E})$. Under Algorithm 1, the estimate $\pi_i(k)=s_i(k)/w_i(k)$ of each node $i$ will converge to the average value $\alpha={\sum_{j=1}^{N}x_j^0}/N$. More specifically, the convergence rate of Algorithm 1 is at least $\gamma= (1-\varepsilon^{N-1}) ^{\frac{1}{N-1}} \in (0, \, 1)$, meaning that there exists a positive constant $C$ such that $\big\|\boldsymbol{\pi}(k) - \alpha \mathbf{1} \big\| \leq C \gamma ^k$ holds for all $k$.
\end{Theorem 1}

{\it Proof}: Combining $\mathbf{w}(K+1)=\mathbf{1}$ in (\ref{Algorithm_I_first_half}) with (\ref{Algorithm_I_second_half}), we obtain
\begin{equation}\label{Theorem_1_second_half}
\left\lbrace \begin{aligned}
& \mathbf{s}(K+l+1) = \mathbf{\Phi}_s(K+l:K+1) \mathbf{s}(K+1)\\
& \mathbf{w}(K+l+1) = \mathbf{\Phi}_w(K+l:K+1) \mathbf{1}\\
\end{aligned} \right.
\end{equation}
for $l \geq 1$. Since for $k \geq K+1$, $\mathbf{P}_{s}(k)$ satisfies the Assumptions 2, 3(a), 4, and 5 in \cite{nedic2010constrained}, according to its Proposition 1(b), the transition matrix $\mathbf{\Phi}_s(K+l:K+1)$ converges to a stochastic vector $\boldsymbol{\varphi}(K+l)$ with a geometric rate with respect to $i$ and $j$, i.e., for all $i, j=1,\ldots,N$ and $l \geq 1$
\begin{equation}\label{Phi_i_j}
\begin{aligned}
\Big|[\mathbf{\Phi}_s(K+l:K+1)]_{ij}-\varphi_i(K+l)\Big| \leq C_0 \gamma^{l-1}
\end{aligned}
\end{equation}
holds with $C_0=2 ({1+\varepsilon^{-N+1}})/({1-\varepsilon^{N-1}})$ and $\gamma= (1-\varepsilon^{N-1}) ^{\frac{1}{N-1}}$. Defining $\mathbf{M}(K+l:K+1)$ as
\begin{equation}\label{D_matrix}
\begin{aligned}
\mathbf{M}(K+l:K+1) \triangleq \mathbf{\Phi}_s(K+l:K+1)-\boldsymbol{\varphi}(K+l) \, \mathbf{1}^T
\end{aligned}
\end{equation}
we have
\begin{equation}\label{D_i_j}
\begin{aligned}
\left|\left[\mathbf{M}(K+l:K+1)\right]_{ij}\right| \leq C_0 \gamma^{l-1}
\end{aligned}
\end{equation}
for all $i, j=1,\ldots,N$ and $l \geq 1$.

Given $\mathbf{P}_{s}(k) = \mathbf{P}_{w}(k)$ for $k \geq K+1$, $\mathbf{\Phi}_s(K+l:K+1)=\mathbf{\Phi}_w(K+l:K+1)$ holds for $l \geq 1$. Further combining (\ref{D_matrix}) with (\ref{Theorem_1_second_half}) leads to
\begin{equation}\label{Theorem_1_second_half_new}
\left\lbrace \begin{aligned}
\mathbf{s}(K+l+1)= & \mathbf{M}(K+l:K+1) \mathbf{s}(K+1) \\
 & + \, \boldsymbol{\varphi}(K+l) \, \mathbf{1}^T \mathbf{s}(K+1) \\
\mathbf{w}(K+l+1)= & \mathbf{M}(K+l:K+1) \mathbf{1} + N \boldsymbol{\varphi}(K+l) \\
\end{aligned} \right.
\end{equation}

Note that $\mathbf{\Phi}_s(K:0)$ is column stochastic since $\mathbf{P}_{s}(k)$ is column stochastic. So from (\ref{Algorithm_I_first_half}), we have a mass conservation property for $\mathbf{s}(k)$ as follows:
\begin{equation}\label{mass_conservation_property}
\begin{aligned}
\mathbf{1}^T \mathbf{s}(K+1) =\mathbf{1}^T  \mathbf{\Phi}_s(K:0) \mathbf{s}(0) =\mathbf{1}^T \mathbf{s}(0)\\
\end{aligned}
\end{equation}
which further leads to
\begin{equation}\label{alpha_new}
\begin{aligned}
\alpha=\frac{\sum_{j=1}^{N}x_j^0}{N} = \frac{\mathbf{1}^T \mathbf{s}(0)}{N}=\frac{\mathbf{1}^T \mathbf{s}(K+1)}{N}\\
\end{aligned}
\end{equation}

Combining (\ref{Theorem_1_second_half_new}) with (\ref{alpha_new}), we have
\begin{equation}\label{pi_alpha}
\begin{aligned}
& \pi_i(K+l+1)-\alpha \\
= & \frac{s_i(K+l+1)}{w_i(K+l+1)}-\frac{\mathbf{1}^T \mathbf{s}(K+1)}{N} \\
= & \frac{s_i(K+l+1)}{w_i(K+l+1)} - \frac{\mathbf{1}^T \mathbf{s}(K+1) w_i(K+l+1)}{N w_i(K+l+1)}\\
= & \frac{[\mathbf{M}(K+l:K+1) \mathbf{s}(K+1)]_i + \varphi_i(K+l) \mathbf{1}^T \mathbf{s}(K+1)} {w_i(K+l+1)}\\
& \, - \frac{\mathbf{1}^T \mathbf{s}(K+1) \big( [\mathbf{M}(K+l:K+1) \mathbf{1}]_i + N \varphi_i(K+l) \big) }{N w_i(K+l+1)}\\
= & \frac{[\mathbf{M}(K+l:K+1) \mathbf{s}(K+1)]_i} {w_i(K+l+1)} \\
& \, - \frac{\mathbf{1}^T \mathbf{s}(K+1) [\mathbf{M}(K+l:K+1) \mathbf{1}]_i} {N w_i(K+l+1)}\\
\end{aligned}
\end{equation}
Therefore, for $i=1,\ldots,N$ and $l \geq 1$, one can obtain
\begin{equation}\label{pi_alpha_absolute}
\begin{aligned}
& \big| \pi_i(K+l+1)-\alpha \big| \\
\leq &  \frac{\big| [\mathbf{M}(K+l:K+1) \mathbf{s}(K+1)]_i \big|} {w_i(K+l+1)} \\
& \, + \frac{\big| \mathbf{1}^T \mathbf{s}(K+1) [\mathbf{M}(K+l:K+1) \mathbf{1}]_i \big|} {N w_i(K+l+1)}\\
\leq & \frac{1}{\varepsilon^N} \big(\max_{j}\big|[\mathbf{M}(K+l:K+1)]_{ij} \big|\big) \big\|\mathbf{s}(K+1)\big\|_1 \\
& \, + \frac{1}{\varepsilon^N} \big|\mathbf{1}^T \mathbf{s}(K+1)\big|  \big(\max_{j}\big|[\mathbf{M}(K+l:K+1)]_{ij} \big|\big)\\
\end{aligned}
\end{equation}
where we used $w_i(K+l+1) \geq \varepsilon^N$ from Lemma 1 in the derivation. Further taking into account the relationship $\big|\mathbf{1}^T \mathbf{s}(K+1)\big| \leq \big\|\mathbf{s}(K+1)\big\|_1$ and (\ref{D_i_j}), we have
\begin{equation}\label{pi_alpha_absolute_new}
\begin{aligned}
\big| \pi_i(K+l+1)-\alpha \big| \leq {2 C_0 \big\|\mathbf{s}(K+1)\big\|_1} {\varepsilon^{-N}} \gamma^{l-1}
\end{aligned}
\end{equation}
for $l\geq 1$.

From (\ref{pi_alpha_absolute_new}), we obtain for $l\geq 1$
\begin{equation}\label{pi_alpha_vector}
\begin{aligned}
\big\|\boldsymbol{\pi}(K+l+1) - \alpha \mathbf{1} \big\| & \leq {2 \sqrt{N} C_0 \big\|\mathbf{s}(K+1)\big\|_1} {\varepsilon^{-N}} \gamma^{l-1} \\
& = C_1 \gamma^{K+l+1}
\end{aligned}
\end{equation}
where $C_1$ is given by
\begin{equation}\label{C_1}
\begin{aligned}
C_1= {2 \sqrt{N} C_0 \big\|\mathbf{s}(K+1)\big\|_1} {\varepsilon^{-N} \gamma^{-K-2}}
\end{aligned}
\end{equation}
Therefore, we have $\big\|\boldsymbol{\pi}(k) - \alpha \mathbf{1} \big\| \leq C_1 \gamma^{k}$ for $k \geq K+2$.

For $k \leq K+1$, given $w_i(k)=1$ and $\mathbf{1}^T \mathbf{s}(k)=\mathbf{1}^T \mathbf{s}(0)$, one can easily obtain
\begin{equation}\label{pi_alpha_K+1}
\begin{aligned}
\pi_i(k)-\alpha =  \frac{s_i(k)}{w_i(k)}-\frac{\mathbf{1}^T \mathbf{s}(k)}{N} =s_i(k)-{\sum_{j=1}^{N}s_j(k)}/{N}\\
\end{aligned}
\end{equation}
Thus, it follows for $k \leq K+1$
\begin{equation}\label{pi_alpha__vector_K+1}
\begin{aligned}
\big\|\boldsymbol{\pi}(k) - \alpha \mathbf{1} \big\| & = \Big({ \sum_{i=1}^{N} \Big( {s_i(k) - {\sum_{j=1}^{N}s_j(k)}/{N}} \Big)^{2} }\Big)^{1/2}\\
& =  \Big(\sum_{i=1}^{N} {\big(s_i(k)\big)^{2}} - {\big(\sum_{j=1}^{N}s_j(k)\big)^{2}/{N}}\Big)^{1/2} \\
& \leq \big\|\mathbf{s}(k) \big\|
\end{aligned}
\end{equation}

Defining $C$ as
\begin{equation}\label{C}
\begin{aligned}
C \triangleq \max \big\{ C_1, \big\|\mathbf{s}(0) \big\|,  \gamma^{-1} \big\|\mathbf{s}(1) \big\|, \ldots, \gamma^{-K-1} {\big\|\mathbf{s}(K+1) \big\|} \big\}
\end{aligned}
\end{equation}
we have
\begin{equation}\label{pi_alpha__vector_all}
\begin{aligned}
\big\|\boldsymbol{\pi}(k) - \alpha \mathbf{1} \big\| \leq C \gamma^{k}
\end{aligned}
\end{equation}
for all $k$. Therefore, each node $i$ will converge to the average value $\alpha={\sum_{j=1}^{N}x_j^0}/N$ with convergence rate at least $\gamma= (1-\varepsilon^{N-1}) ^{\frac{1}{N-1}} \in (0, \, 1)$.
\hfill{$\blacksquare$}

From Theorem 1, it can be seen that a smaller $\gamma$ leads to a faster convergence. Given the relationship $\gamma= (1-\varepsilon^{N-1}) ^{\frac{1}{N-1}}$, to get a faster convergence speed, i.e., a smaller $\gamma$, it suffices to increase $\varepsilon$, which amounts to reducing the size of the range $(\varepsilon, \, 1)$ for the random selection of coupling weights $p_{ji}^s(k)$ and $p_{ji}^w(k)$ with $k \geq K+1$. Note that although the reduced range $(\varepsilon, \, 1)$ enables an honest-but-curious node to get a better range estimation of node $i$'s intermediate states $s_i(k)$ and $w_i(k)$ for $k \geq K+1$ from $p_{ji}^s(k)s_i(k)$ and $p_{ji}^w(k)w_i(k)$, it does not affect the privacy protection of node $i$'s initial state $x_i^0$, as will be shown in our privacy analysis in the following subsection. It is also worth noting that to satisfy the requirements of random weights selection in Algorithm 1, $\varepsilon$ cannot be arbitrarily close to $1$. In fact, $\varepsilon$ must satisfy $\varepsilon < 1/(\max_i{D_i^{out}}+1)$.

\subsection{Privacy-Preserving Performance Analysis}

In this subsection, we rigorously prove that Algorithm 1 is able to achieve the privacy defined in Definition \ref{privacy_preservation_definition}. We consider two different scenarios: 1) a single adversary acting on its own (i.e., without collusion with other adversaries) and 2) multiple adversaries colluding with each other.

\subsubsection{Single honest-but-curious node case} In this situation, we make the following assumption.
\begin{Assumption 1}
	We assume that there are $1 \leq Q \leq N$ honest-but-curious nodes which try to infer node $i$'s initial value without sharing information with each other.
\end{Assumption 1}

\begin{Theorem 2}
	Consider a distributed network of $N$ nodes interacting on a strongly connected graph $\mathcal{G}=(\mathcal{V}, \, \mathcal{E})$. Under Assumption 1, Algorithm 1 can preserve the privacy of node $i$ if $|\mathcal{N}_i^{out} \cup \mathcal{N}_i^{in}| \geq 2$ holds.
\end{Theorem 2}

{\it Proof}: To show that the privacy of node $i$ can be preserved, we have to show that the privacy of the initial value $x_i^0$ of node $i$ can be preserved against any honest-but-curious node $j$,  i.e., node $j$ cannot estimate the value of $x_i^0$ with any accuracy. Our idea is based on the indistinguishability of $x_i^0$'s {\it arbitrary} variation to node $j$. According to Algorithm 1 and Assumption 1, all the information accessible to node $j$, denoted as $\mathcal{I}_j$, are given as follows:
\begin{equation}\label{collected_information}
\begin{aligned}
\mathcal{I}_j= & \big\{ \mathcal{I}_j^{\text{state}}(k) \cup \mathcal{I}_j^{\text{send}}(k) \cup \mathcal{I}_j^{\text{receive}}(k) \, \big| \, k=0,1,\ldots \big\}\\
& \cup \big\{ w_m(k)=1 \, \big| \, m \in \mathcal{V}, \ k=0,1,\ldots,K+1 \big\} \\
\end{aligned}
\end{equation}
where
\begin{equation}\label{collected_information_detail}
\begin{aligned}
& \mathcal{I}_j^{\text{state}}(k) = \big\{s_j(k), w_j(k) \big\}\\
& \mathcal{I}_j^{\text{send}}(k) = \big\{ p_{mj}^s(k) s_j(k), p_{mj}^w(k) w_j(k) \ | \ m \in \mathcal{N}_j^{out} \cup \{j\} \big\}\\
& \mathcal{I}_j^{\text{receive}}(k) = \big\{ p_{jn}^s(k) s_n(k), p_{jn}^w(k) w_n(k) \ | \ n \in \mathcal{N}_j^{in} \big\}\\
\end{aligned}
\end{equation}
represent the respective state information, sent information, and received information of node $j$ at iteration $k$.

The only information available to node $j$ to infer $x_i^0$ is $\mathcal{I}_j$. So if we can show that under {\it any initial value} $\tilde{x}_i^0 \neq x_i^0$, the information accessible to node $j$, i.e., $\tilde{\mathcal{I}}_j$, could be exactly the same as $\mathcal{I}_j$ in (\ref{collected_information}) and (\ref{collected_information_detail}), then node $j$ has no way to even find a range for the initial value $x_i^0$. Therefore, we only need to prove that for any $\tilde{x}_i^0 \neq x_i^0$, $\tilde{\mathcal{I}}_j = \mathcal{I}_j$ could hold.

Under the condition $|\mathcal{N}_i^{out} \cup \mathcal{N}_i^{in}| \geq 2$, there must exist a node $l \in \mathcal{N}_i^{out} \cup \mathcal{N}_i^{in}$ such that  $l \neq j$ holds. Next we show that there exist initial values of $x_l^0$ and coupling weights satisfying the requirements in Algorithm 1 that make $\tilde{\mathcal{I}}_j = \mathcal{I}_j$ hold for any $\tilde{x}_i^0 \neq x_i^0$. More specifically, under the initial condition $\tilde{x}_l^0=x_i^0+x_l^0-\tilde{x}_i^0$, we consider two cases, $l \in \mathcal{N}_i^{out}$ and $l \in \mathcal{N}_i^{in}$, respectively (note that if $l \in \mathcal{N}_i^{out} \cap \mathcal{N}_i^{in}$ holds, either case can be used in the argument to draw a same conclusion):

Case I: If $l \in \mathcal{N}_i^{out}$ holds, we can easily obtain $\tilde{\mathcal{I}}_j = \mathcal{I}_j$ for any $\tilde{x}_i^0 \neq x_i^0$ under the following coupling weights
\begin{equation}\label{proper_coupling_weights_I}
\left\lbrace\begin{aligned}
& \tilde{p}_{mn}^s(0)=p_{mn}^s(0) \quad \forall \, m \in \mathcal{V}, \, n\in \mathcal{V}\setminus \{i, \, l\} \\
& \tilde{p}_{mi}^s(0)=p_{mi}^s(0) x_i^0/\tilde{x}_i^0 \quad \forall \, m \in \mathcal{V}\setminus \{l\} \\
& \tilde{p}_{li}^s(0)=(p_{li}^s(0) x_i^0 + \tilde{x}_i^0- x_i^0 )/\tilde{x}_i^0\\
& \tilde{p}_{ml}^s(0)=p_{ml}^s(0) x_l^0/(x_i^0+x_l^0-\tilde{x}_i^0) \quad \forall \, m \in \mathcal{V}\setminus \{l\} \\
& \tilde{p}_{ll}^s(0)=(p_{ll}^s(0) x_l^0 - \tilde{x}_i^0 + x_i^0 )/(x_i^0+x_l^0-\tilde{x}_i^0)\\
& \tilde{p}_{mn}^s(k)=p_{mn}^s(k) \quad \forall \, m,n \in \mathcal{V}, \, k=1,2,\ldots\\
& \tilde{p}_{mn}^w(k)=p_{mn}^w(k) \quad \forall \, m,n \in \mathcal{V}, \, k=0,1,\ldots\\
\end{aligned}\right.
\end{equation}
where ``$\setminus$'' represents set subtraction.

Case II: If $l \in \mathcal{N}_i^{in}$ holds, it can be easily verified that $\tilde{\mathcal{I}}_j = \mathcal{I}_j$ is true for any $\tilde{x}_i^0 \neq x_i^0$ under the following coupling weights

\begin{equation}\label{proper_coupling_weights_II}
\left\lbrace\begin{aligned}
& \tilde{p}_{mn}^s(0)=p_{mn}^s(0) \quad \forall \, m \in \mathcal{V}, \, n\in \mathcal{V}\setminus \{i, \, l\}\\
& \tilde{p}_{mi}^s(0)=p_{mi}^s(0) x_i^0/\tilde{x}_i^0 \quad \forall \, m \in \mathcal{V}\setminus \{i\} \\
& \tilde{p}_{ii}^s(0)=(p_{ii}^s(0) x_i^0 + \tilde{x}_i^0- x_i^0 )/\tilde{x}_i^0\\
& \tilde{p}_{ml}^s(0)=p_{ml}^s(0) x_l^0/(x_i^0+x_l^0-\tilde{x}_i^0) \ \ \forall \, m \in \mathcal{V}\setminus \{i\} \\
& \tilde{p}_{il}^s(0)=(p_{il}^s(0) x_l^0 - \tilde{x}_i^0 + x_i^0 )/(x_i^0+x_l^0-\tilde{x}_i^0)\\
& \tilde{p}_{mn}^s(k)=p_{mn}^s(k)  \quad \forall \, m,n \in \mathcal{V}, \, k=1,2,\ldots\\
& \tilde{p}_{mn}^w(k)=p_{mn}^w(k) \quad \forall \, m,n \in \mathcal{V}, \, k=0,1,\ldots\\
\end{aligned}\right.
\end{equation}

Summarizing cases I and II, we have $\tilde{\mathcal{I}}_j = \mathcal{I}_j$ for any $\tilde{x}_i^0 \neq x_i^0$, which means that node $j$ cannot estimate the value of $x_i^0$ with any accuracy based on its accessible information. Therefore, under Algorithm 1 and Assumption 1, the privacy of node $i$ can be preserved if $|\mathcal{N}_i^{out} \cup \mathcal{N}_i^{in}| \geq 2$ holds. \hfill{$\blacksquare$}

Next we show that if the conditions in Theorem 2 cannot be met, then the privacy of node $i$ can be breached. 

\begin{Theorem 3}
	Consider a distributed network of $N$ nodes interacting on a strongly connected graph $\mathcal{G}=(\mathcal{V}, \, \mathcal{E})$. Under Algorithm 1 and Assumption 1, the privacy of node $i$ cannot be preserved against node $j$ if node $j$ is the only in-neighbor and out-neighbor of node $i$, i.e., $\mathcal{N}_i^{out} = \mathcal{N}_i^{in}= \{j\}$. In fact, under $\mathcal{N}_i^{out} = \mathcal{N}_i^{in}= \{j\}$, the initial value $x_i^0$ of node $i$ can be uniquely determined by node $j$.
\end{Theorem 3}

{\it Proof}: Since $\mathcal{N}_i^{out} = \mathcal{N}_i^{in}= \{j\}$ holds, one can get the dynamics of $s_i$ and $w_i$ from (\ref{Algorithm_I_s_w_update}) as follows:
\begin{equation}\label{node_i_s_w}
\left\lbrace \begin{aligned}
& s_i(k+1) = p_{ii}^s(k)s_i(k) + p_{ij}^s(k)s_j(k)\\
& w_i(k+1) = p_{ii}^w(k)w_i(k) + p_{ij}^w(k)w_j(k)
\end{aligned}\right.
\end{equation}
Under the requirements of coupling weights in Algorithm 1, we have $p_{ii}^s(k) + p_{ji}^s(k)=1$ and $p_{ii}^w(k) + p_{ji}^w(k)=1$. So it follows
\begin{equation}\label{node_i_s_w_2}
\left\lbrace \begin{aligned}
& s_i(k) = p_{ii}^s(k)s_i(k) + p_{ji}^s(k)s_i(k)\\
& w_i(k) = p_{ii}^w(k)w_i(k) + p_{ji}^w(k)w_i(k)
\end{aligned}\right.
\end{equation}
Combining (\ref{node_i_s_w}) and (\ref{node_i_s_w_2}) leads to
\begin{equation}\label{node_i_s_w_3}
\left\lbrace \begin{aligned}
& s_i(k+1)- s_i(k)= p_{ij}^s(k)s_j(k) - p_{ji}^s(k)s_i(k)\\
& w_i(k+1)- w_i(k)= p_{ij}^w(k)w_j(k) - p_{ji}^w(k)w_i(k)
\end{aligned}\right.
\end{equation}
and further
\begin{equation}\label{node_i_s_w_4}
\left\lbrace \begin{aligned}
& s_i(k)- s_i(0)=\sum\limits_{l=0}^{k-1} \left[ p_{ij}^s(l)s_j(l) - p_{ji}^s(l)s_i(l) \right]\\
& w_i(k)- w_i(0)=\sum\limits_{l=0}^{k-1} \left[ p_{ij}^w(l)w_j(l) - p_{ji}^w(l)w_i(l)
\right]
\end{aligned}\right.
\end{equation}
Note that the right-hand side of (\ref{node_i_s_w_4}) is accessible to the honest-but-curious node $j$ because $p_{ij}^s(l)s_j(l)$ and $p_{ij}^w(l)w_j(l)$ are computed and sent by node $j$, and $p_{ji}^s(l)s_i(l)$ and $p_{ji}^w(l)w_i(l)$ are received by node $j$. Further taking into consideration $w_i(0)=1$, node $j$ can infer $w_i(k)$ for all $k$. Given $p_{ji}^s(k)=p_{ji}^w(k)$ for $k \geq K+1$, node $j$ can also infer $s_i(k)$ below for $k\geq K+1$.  
\begin{equation}\label{node_i_s_5}
s_i(k)=\frac{p_{ji}^s(k)s_i(k)}{p_{ji}^w(k)w_i(k)} w_i(k)
\end{equation}
Therefore, we have that node $j$ can uniquely infer $x_i^0=s_i(0)$ based on (\ref{node_i_s_w_4}). \hfill{$\blacksquare$}

{\subsubsection{Multiple honest-but-curious nodes colluding with each other} In this situation, we make the following assumption.}
\begin{Assumption 2}
	We assume that a set of honest-but-curious nodes $\mathcal{A}$ share information with each other to infer the initial value $x_i^0$ of node $i \notin \mathcal{A}$.
\end{Assumption 2}

\begin{Theorem 4}
	Consider a network of $N$ nodes interacting on a strongly connected graph $\mathcal{G}=(\mathcal{V}, \, \mathcal{E})$. Under Algorithm 1 and Assumption 2, the privacy of node $i$ can be preserved against the set of honest-but-curious nodes $\mathcal{A}$ if $(\mathcal{N}_i^{out} \cup \mathcal{N}_i^{in}) \not\subset \mathcal{A}$ holds, i.e., there exits at least one node that belongs to $\mathcal{N}_i^{out} \cup \mathcal{N}_i^{in}$ but not $\mathcal{A}$.
\end{Theorem 4}

{\it Proof}: To show that the privacy of node $i$ can be preserved, we have to show that no honest-but-curious node $j\in \mathcal{A}$ can estimate the value of $x_i^0$ with any accuracy. According to Assumption 2, each node in $\mathcal{A}$ has access to the information accessible to any node in $\mathcal{A}$. So we represent the accessible information as
\begin{equation}\label{collected_information_A}
\begin{aligned}
\mathcal{I_A}= \big\{ \mathcal{I}_j \, \big| \, j\in \mathcal{A} \big\}\\
\end{aligned}
\end{equation}
where $\mathcal{I}_j$ is given by (\ref{collected_information}). Following the same line of reasoning in Theorem 2, to prove that node $j\in \mathcal{A}$ cannot estimate the value of $x_i^0$ with any accuracy, it suffices to prove that for any $\tilde{x}_i^0 \neq x_i^0$, the information accessible to node $j$, i.e., $\tilde{\mathcal{I}}_\mathcal{A}$, could be exactly the same as $\mathcal{I_A}$.

If $(\mathcal{N}_i^{out} \cup \mathcal{N}_i^{in}) \not\subset \mathcal{A}$ is true, then there must exist a node $l \in \mathcal{N}_i^{out} \cup \mathcal{N}_i^{in}$ such that $l \notin \mathcal{A}$ holds. It can be easily verified that under the initial condition $\tilde{x}_l^0=x_i^0+x_l^0-\tilde{x}_i^0$, there exist respective coupling weights in (\ref{proper_coupling_weights_I}) and (\ref{proper_coupling_weights_II}) for $l \in \mathcal{N}_i^{out}$ and $l \in \mathcal{N}_i^{in}$ that satisfy the requirements in Algorithm 1 and make $\tilde{\mathcal{I}}_\mathcal{A}=\mathcal{I_A}$ hold under any $\tilde{x}_i^0 \neq x_i^0$. 

Thus, we can have $\tilde{\mathcal{I}}_\mathcal{A}=\mathcal{I_A}$ for any $\tilde{x}_i^0 \neq x_i^0$, meaning that node $j \in \mathcal{A}$ cannot estimate the value of $x_i^0$ with any accuracy even based on the information accessible to the entire set $\mathcal{A}$. Therefore, under Algorithm 1 and Assumption 2, the privacy of node $i$ can be preserved against the set of honest-but-curious nodes $\mathcal{A}$ if $(\mathcal{N}_i^{out} \cup \mathcal{N}_i^{in}) \not\subset \mathcal{A}$ holds. \hfill{$\blacksquare$}

Next we show that if the conditions in Theorem 4 cannot be met, then the privacy of node $i$ can be breached. 

\begin{Theorem 5}
	Consider a distributed network of $N$ nodes interacting on a strongly connected graph $\mathcal{G}=(\mathcal{V}, \, \mathcal{E})$. Under Algorithm 1 and Assumption 2, the privacy of node $i$ cannot be preserved when all the in-neighbors and out-neighbors belong to $\mathcal{A}$, i.e., $(\mathcal{N}_i^{out} \cup \mathcal{N}_i^{in}) \subset \mathcal{A}$. In fact, when $(\mathcal{N}_i^{out} \cup \mathcal{N}_i^{in}) \subset \mathcal{A}$ is true, the initial value $x_i^0$ of node $i$ can be uniquely determined by honest-but-curious nodes $j\in \mathcal{A}$.
\end{Theorem 5}

{\it Proof}: From (\ref{Algorithm_I_s_w_update}), we have the dynamics of $s_i$ and $w_i$ as follows:
\begin{equation}\label{node_i_s_w_A}
\left\lbrace \begin{aligned}
& s_i(k+1) = \sum\limits_{n\in \mathcal{N}_i^{in}}{p_{in}^s(k)s_n(k)} + p_{ii}^s(k)s_i(k)\\
& w_i(k+1) = \sum\limits_{n\in \mathcal{N}_i^{in}}{p_{in}^w(k)w_n(k)} + p_{ii}^w(k)w_i(k)
\end{aligned}\right.
\end{equation}
Under the requirements of coupling weights in Algorithm 1, we have $p_{ii}^s(k) + \sum_{m\in \mathcal{N}_i^{out}}{p_{mi}^s(k)}=1$ and $p_{ii}^w(k) + \sum_{m\in \mathcal{N}_i^{out}}{p_{mi}^w(k)}=1$. Plugging these equations into (\ref{node_i_s_w_A}), we can obtain
\begin{equation}\label{node_i_s_w_3_A}
\left\lbrace \begin{aligned}
& s_i(k+1)- s_i(k) = \\
& \qquad \sum\limits_{n\in \mathcal{N}_i^{in}}{p_{in}^s(k)s_n(k)} - \sum\limits_{m\in \mathcal{N}_i^{out}}{p_{mi}^s(k)s_i(k)}\\
& w_i(k+1)- w_i(k)= \\
& \qquad \sum\limits_{n\in \mathcal{N}_i^{in}}{p_{in}^w(k)w_n(k)} - \sum\limits_{m\in \mathcal{N}_i^{out}}{p_{mi}^w(k)w_i(k)}\\
\end{aligned}\right.
\end{equation}
and further
\begin{equation}\label{node_i_s_w_4_A}
\left\lbrace \begin{aligned}
& s_i(k)- s_i(0)= \\
& \qquad \sum\limits_{l=0}^{k-1} \Big[ \sum\limits_{n\in \mathcal{N}_i^{in}}{p_{in}^s(l)s_n(l)} - \sum\limits_{m\in \mathcal{N}_i^{out}}{p_{mi}^s(l)s_i(l)} \Big]\\
& w_i(k)- w_i(0)= \\
& \qquad \sum\limits_{l=0}^{k-1} \Big[ \sum\limits_{n\in \mathcal{N}_i^{in}}{p_{in}^w(l)w_n(l)} - \sum\limits_{m\in \mathcal{N}_i^{out}}{p_{mi}^w(l)w_i(l)} \Big]
\end{aligned}\right.
\end{equation}
Note that under Assumption 2, each honest-but-curious node $j \in \mathcal{A}$ has access to the information $\mathcal{I_A}$ in (\ref{collected_information_A}). If $(\mathcal{N}_i^{out} \cup \mathcal{N}_i^{in}) \subset \mathcal{A}$ is true, then all terms in the right-hand side of (\ref{node_i_s_w_4_A}) belong to $\mathcal{I_A}$, and hence are accessible to every node $j \in \mathcal{A}$. By further taking into consideration $w_i(0)=1$, node $j$ can uniquely infer $w_i(k)$ for all $k$. 

Given $p_{li}^s(k)=p_{li}^w(k)$ for $l \in \mathcal{N}_i^{out} \subset \mathcal{A}$ and $k \geq K+1$, each node $j\in \mathcal{A}$ can infer $s_i(k)$ for $k\geq K+1$ by using the following relationship
\begin{equation}\label{node_i_s_5_A}
s_i(k)=\frac{p_{li}^s(k)s_i(k)}{p_{li}^w(k)w_i(k)} w_i(k) \quad \text{for}  \quad  k \geq K+1
\end{equation}
Note that $p_{li}^s(k)s_i(k)$ and $p_{li}^w(k)w_i(k)$ are known to node $j$ since $l \in \mathcal{A}$ holds. Further making use of (\ref{node_i_s_w_4_A}), each node $j\in \mathcal{A}$ can uniquely determine the value of $x_i^0=s_i(0)$. \hfill{$\blacksquare$}

\begin{Remark 1}\label{collude}
	Theorems 4 and 5 are obtained under Assumption 2 that honest-but-curious nodes in set $\mathcal{A}$ collude with each other. If there are more than one set of honest-but-curious nodes in a network and these sets do not collude with each other, then following a similar argument as in Theorems 4 and 5, we can prove that the privacy of node $i$ can be preserved if $\mathcal{N}_i^{out} \cup \mathcal{N}_i^{in}$ is not included in any set of colluding honest-but-curious nodes.
\end{Remark 1}

\begin{Remark 2}\label{tradeoff}
	From the above analysis, we know that besides the given topology conditions, introducing $\mathbf{P}_{s}(k) \neq \mathbf{P}_{w}(k)$ for $k\leq K$ is key to protect privacy against honest-but-curious nodes. However, making $\mathbf{P}_{s}(k) \neq \mathbf{P}_{w}(k)$ delays the convergence and hence leads to a trade-off between privacy preservation and convergence speed. This is confirmed in our numerical simulations in Fig. \ref{convergence_error_figure} which showed that convergence only initiated after $k = K+1$.
\end{Remark 2}

\begin{Remark 3}\label{tradeoff2}	
	If an adversary can obtain side information, then a larger $K$ protects the privacy of more intermediate states $s_i(k)$ for $1\leq k\leq K$. This is because for $k \geq K+1$, $\pi(k)$ can be easily obtained by an adversary due to $\mathbf{P}_{s}(k) = \mathbf{P}_{w}(k)$ for $k\geq K+1$, which makes states $s_i(k)$ for $k \geq K+1$ easier to infer through $s_i(k)=\pi(k)w_i(k)$ if side information about $w_i(k)$ is available. Therefore, although a smaller $K$ leads to a faster convergence, as discussed in Remark 2, a larger $K$ can protect more intermediate states ($s_i(k)$ for $1\leq k \leq K$) when an adversary can obtain side information. Since our paper focuses on the privacy preservation of the initial state $x_i^0 = s_i(0)$, we can set $K$ to be any non-negative integer.
\end{Remark 3}

\section{Extended Privacy-Preserving Average Consensus Against Eavesdroppers}

It is well known that exchanging information in the unencrypted plaintext form is vulnerable to eavesdroppers who steal information by wiretapping communication links. So in this section, we extend our proposed Algorithm 1 to provide protection against eavesdroppers by employing partially homomorphic encryption. To this end, we first introduce the definition of eavesdroppers as follows.

\begin{Definition 4}
	An eavesdropper is an external attacker who knows the network interaction dynamics and topology, and is able to wiretap all communication links and intercept all exchanged messages.
\end{Definition 4}

Generally speaking, an eavesdropper is more disruptive than an honest-but-curious node in terms of information breaches because it can access all messages exchanged in the network whereas the latter can only access the messages destined to it. However, an honest-but-curious node does have one piece of information that is unknown to an external eavesdropper, i.e., the honest-but-curious node $i$ has access to its internal initial value $x_i^0$. It is also worth noting that existing correlated-noise based privacy-preserving average consensus approaches in \cite{kefayati2007secure, manitara2013privacy, he2016private, mo2017privacy} fail to provide protection against eavesdroppers modeled in this paper.

To preserve privacy against eavesdroppers, we employ the Paillier cryptosystem in Section II-C to encrypt exchanged messages. So each node $i$ needs to generate its public key $k_i^p=(h_i, \, g_i)$ and private key $k_i^s=(\lambda_i, \, \mu_i)$ before the iterative consensus starts. It is worth noting that under the partially homomorphic cryptography framework, every node can have access to every other nodes' public key (using e.g., flooding \cite{tanenbaum2011computer}) without the assistance of any third party or data aggregator, even when the communication links are directed. This is in disparate difference from commonly used encryption schemes which rely on either bi-directional communications or a trusted third party to manage keys \cite{goldreich2009foundations}.

\noindent\rule{0.49\textwidth}{0.5pt}
\noindent\textbf{Algorithm 2:} Extended privacy-preserving average consensus algorithm against eavesdroppers

\vspace{-0.2cm}\noindent\rule{0.49\textwidth}{0.5pt}
\begin{enumerate}
	\item Positive integer $K$ and parameter $\varepsilon\in (0, \, 1)$ are known to every node. Each node $i$ generates a public key $k_i^p=(h_i, \, g_i)$ and a private key $k_i^s=(\lambda_i, \, \mu_i)$, and initializes $s_i(0)=x_i^0$, $w_i(0)=1$, and $\pi_i(0)=s_i(0)/w_i(0)$.
	\item At iteration $k$,
          \begin{enumerate}
          	\item If $k \leq K$, node $i$ generates one set of random coupling weights $\big\{p_{ji}^s(k) \in \mathbb{R} \, \big| \, j\in \mathcal{N}_i^{out} \cup \{i\} \big\}$ with the sum of this set equal to $1$, and assigns $p_{ii}^w(k)=1$ and $ p_{ji}^w(k)=0$ for $j\in \mathcal{N}_i^{out}$; otherwise, node $i$ generates one set of random coupling weights $\big\{p_{ji}^s(k) \in  (\varepsilon, \, 1) \, \big| \, j\in \mathcal{N}_i^{out} \cup \{i\} \big\}$ satisfying the sum $1$ condition, and assigns $p_{ji}^w(k)=p_{ji}^s(k)$ for $j\in \mathcal{N}_i^{out} \cup \{i\}$.
          	\item Node $i$ computes $p_{ji}^s(k) s_i(k)$ and $p_{ji}^w(k) w_i(k)$ for $j\in \mathcal{N}_i^{out}$, uses the public key $k_j^p$ of node $j$ to encrypt these values, and sends them to node $j$.
          	\item After receiving the encrypted messages $p_{ij}^s(k) s_j(k)$ and $p_{ij}^w(k) w_j(k)$ from its in-neighbors $j \in \mathcal{N}_i^{in}$, node $i$ uses its private key $k_i^s$ to decrypt them, and updates $s_i$ and $w_i$ as follows:
          	\begin{equation}\label{Algorithm_I_s_w_update_2}
          	\left\lbrace \begin{aligned}
          	& s_i(k+1) = \sum_{j\in \mathcal{N}_i^{in}\cup \{i\}} p_{ij}^s(k) s_j(k)\\
          	& w_i(k+1) = \sum_{j\in \mathcal{N}_i^{in}\cup \{i\}} p_{ij}^w(k) w_j(k)
          	\end{aligned} \right.
          	\end{equation}	
          	\item Node $i$ computes the estimate as $\pi_i(k+1)=s_i(k+1)/w_i(k+1)$.
          \end{enumerate}
\end{enumerate}
\vspace{-0.2cm}\rule{0.49\textwidth}{0.5pt}

\section{Results Validation}

We used both numerical simulations and hardware experiments to verify the effectiveness of our proposed approach.

\subsection{Numerical Simulations}

\begin{figure}[h]
	\begin{center}
		\includegraphics[width=0.2\textwidth]{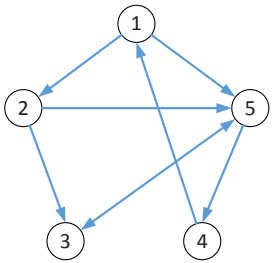}
	\end{center}
	\caption{Illustration of a strongly connected graph with $5$ nodes.}
	\label{graph}
\end{figure}

\begin{figure}[h]
	\begin{center}
		\includegraphics[width=0.45\textwidth]{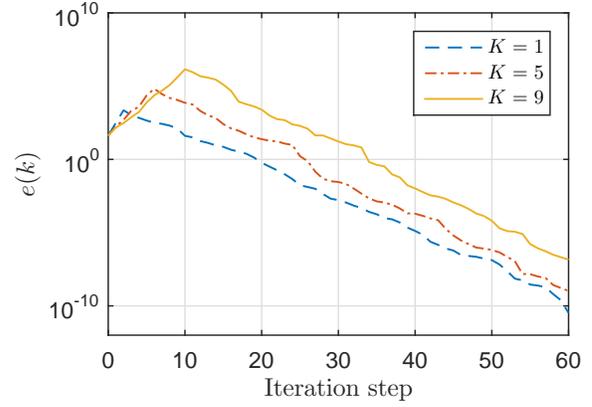}
	\end{center}
	\caption{The evolution of error $e(k)$ under different $K$ for a network of 5 nodes.}
	\label{convergence_error_figure}
\end{figure}

We first evaluated the convergence performance of our proposed Algorithm 1. In this simulation, we considered a network of $N=5$ nodes interacting on a strongly connected graph, which is shown in Fig. \ref{graph}. $\varepsilon$ was set to $0.01$. For iterations $k \leq K$, following the requirements of Algorithm 1, the coupling weights $p_{ij}^s(k)$ for $j\in \mathcal{N}_i^{out} \cup \{i\}$ were chosen from $(-10, \, 10)$. The initial values $x_i^0$ for $i=1, \ldots, N$ were randomly chosen from $(0, \, 50)$. We used $e(k)$ to measure the error between the estimate $\pi_i(k)=s_i(k)/w_i(k)$ and the true average value $\alpha={\sum_{j=1}^{N}x_j^0}/{N}$ at iteration $k$, i.e.,
\begin{equation}\label{convergence_error}
\begin{aligned}
e(k)=\big\|\boldsymbol{\pi}(k) - \alpha \mathbf{1} \big\|= \big( \sum\limits_{i=1}^{N}(\pi_i(k)-\alpha)^{2}\big)^{1/2}
\end{aligned}
\end{equation}
Three experiments were conducted with $K$ being set to $1$, $5$, and $9$, respectively. The evolution of $e(k)$ is shown in Fig. \ref{convergence_error_figure}. It can be seen that $e(k)$ approached to $0$, meaning that every node converged to the average value $\alpha=\sum_{j=1}^{N}x_j^0/{N}$. It is also worth noting that from Fig. \ref{convergence_error_figure}, we can see that Algorithm 1 did not start to converge until iteration step $k \geq K+1$, which confirmed our analysis in Remark \ref{tradeoff}.

We then evaluated the privacy-preserving performance of Algorithm 1. Under the same network setup as in the previous simulation, we considered the case where a set of honest-but-curious nodes $2$, $3$, and $4$ colluded to infer the initial value $x_1^0$ of node $1$. So we have $\mathcal{A}=\{2,\, 3,\, 4\}$. Note that $\mathcal{N}_1^{out} \cup \mathcal{N}_1^{in}=\{2,\, 4,\, 5\} \not\subset \mathcal{A}$ holds for node $1$. Two experiments were conducted with $x_1^0$ being set to $40$ and $-40$, respectively. $x_2^0, \ldots, x_{5}^0$ were randomly chosen from $(0, \, 50)$. $K$ was set to $1$. The other parameters were the same as the first simulation, and the maximal iteration step was $M=100$. To infer the value of $x_1^0$, the nodes in set $\mathcal{A}$ were able to construct the following linear equations based on the accessible information $\mathcal{I_A}$
\begin{equation}\label{node_1_s_w}
\begin{aligned}
s_1(k+1)- s_1(k) + \Delta s_1(k) =p_{14}^s(k)s_4(k) - p_{21}^s(k)s_1(k)
\end{aligned}
\end{equation}
for $k=0,1,\ldots,M$ and
\begin{equation}\label{node_1_s_w_0}
\begin{aligned}
w_1(k+1)- w_1(k) + \Delta w_1(k) = p_{14}^w(k)w_4(k)- p_{21}^w(k)w_1(k)
\end{aligned}
\end{equation}
for $k=K+1,\ldots,M$ where $\Delta s_1(k)$ and $\Delta w_1(k)$ are given by
\begin{equation*}
\begin{aligned}
& \Delta s_1(k)= p_{51}^s(k)s_1(k) \\
& \Delta w_1(k)= p_{51}^w(k)w_1(k) \\
\end{aligned}
\end{equation*}
Given $p_{21}^s(k)=p_{21}^w(k)$ for $k\geq K+1$, set $\mathcal{A}$ also constructed the following relationship
\begin{equation}\label{node_1_s_w_1}
s_1(k)-\pi_1(k) w_1(k)=0
\end{equation}
for $k=K+1,K+2,\ldots,M$ where $\pi_1(k)={p_{21}^s(k)s_1(k)}/{p_{21}^w(k)w_1(k)}$ is known to set $\mathcal{A}$ since both ${p_{21}^s(k)s_1(k)}$ and ${p_{21}^w(k)w_1(k)}$ belong to $\mathcal{I_A}$.

\begin{figure}
	\begin{center}
		\includegraphics[width=0.5\textwidth]{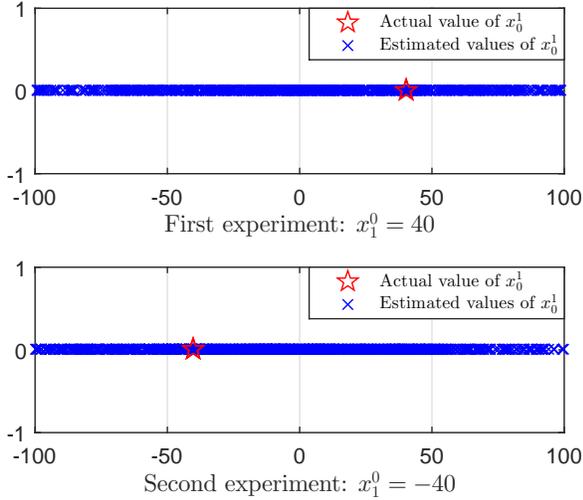}
	\end{center}
	\caption{Estimation of $x_1^0$ by the set of honest-but-curious nodes $2$, $3$, and $4$. In each experiment, the actual value of $x_1^0$ is represented by the star, and the estimated values of $x_1^0$ are represented by the x-marks.}
	\label{Simulation_2_figure}
\end{figure}

The total number of linear equations (\ref{node_1_s_w}), (\ref{node_1_s_w_0}), and (\ref{node_1_s_w_1}) is $3M-2K+1$, and in these equations there are $4M-2K+3$ variables unknown to set $\mathcal{A}$, i.e., $s_1(0), \ldots, s_1(M+1), \Delta s_1(0), \ldots, \Delta s_1(M), w_1(K+2), \ldots, w_1(M+1), \Delta w_1(K+1),$ $\ldots, \Delta w_1(M)$. So there are infinitely many solutions since the number of equations is less than the number of unknown variables. In order to uniquely determine the value of $x_1^0$, we used the least-squares solution to estimate $x_1^0$. In each experiment, set $\mathcal{A}$ estimated $x_1^0$ for $1000$ times, with recorded estimation results shown in Fig. \ref{Simulation_2_figure}. It can be seen that set $\mathcal{A}$ cannot get a good estimate of $x_1^0$.

\begin{figure}
	\begin{center}
		\includegraphics[width=0.45\textwidth]{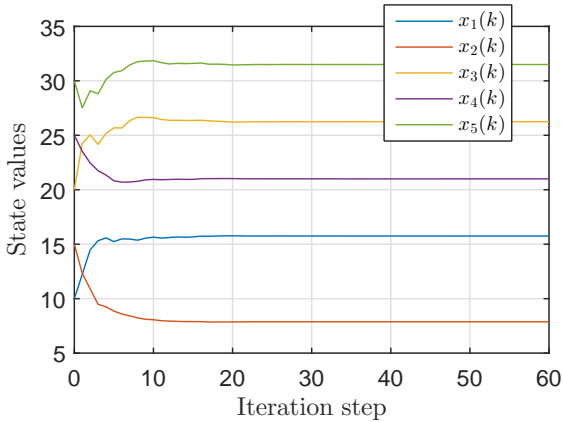}
	\end{center}
	\caption{The evolution of $x_i(k)$ under the approach proposed in \cite{huang2012differentially}.}
	\label{comparison_Zhenqi_Huan_WPES}
\end{figure}

\begin{figure}
	\begin{center}
		\includegraphics[width=0.45\textwidth]{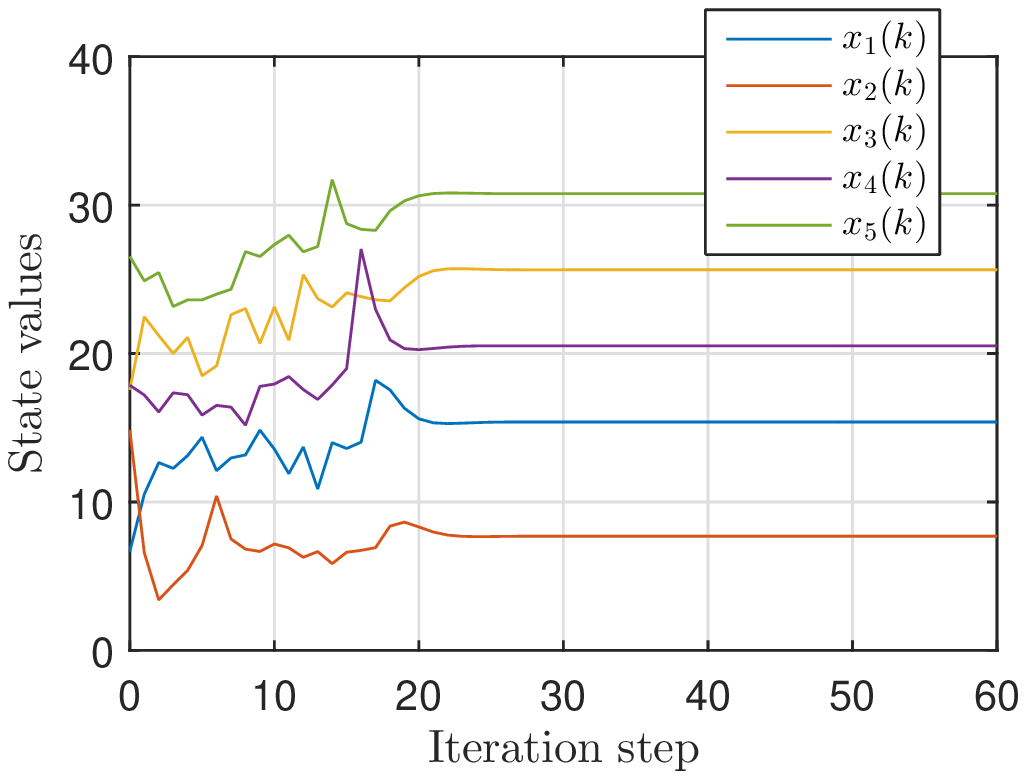}
	\end{center}
	\caption{The evolution of $x_i(k)$ under the approach proposed in \cite{manitara2013privacy}.}
	\label{comparison_Nicolaos_Manitara_ECC}
\end{figure}

\begin{figure}
	\begin{center}
		\includegraphics[width=0.45\textwidth]{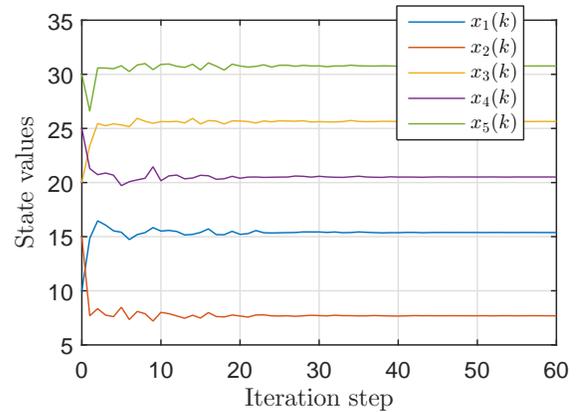}
	\end{center}
	\caption{The evolution of $x_i(k)$ under the approach proposed in \cite{mo2017privacy}.}
	\label{comparison_Yilin_Mo}
\end{figure}

We also compared our Algorithm 1 with existing data-obfuscation based approaches, more specifically, the differential-privacy based approach in \cite{huang2012differentially}, the finite-noise-sequence approach in \cite{manitara2013privacy}, and the decaying-noise approach in \cite{mo2017privacy}. Under the same network setup as in the previous simulations, we set the initial values to $\{10, 15, 20, 25, 30\}$, respectively, with the average value given by $20$. The weight matrix $\mathbf{W}$ was adopted from \cite{huang2012differentially}, i.e., the $ij$-th entry was set to $w_{ij}={1}/{(|\mathcal{N}_j^{out}|+1)}$ for $i\in \mathcal{N}_j^{out} \cup \{j\}$ and $w_{ij} = 0$ for $i\notin \mathcal{N}_j^{out}\cup \{j\}$. Since the graph is directed and imbalanced, and does not satisfy the undirected or balanced assumption in \cite{huang2012differentially, manitara2013privacy, mo2017privacy}, the approaches in \cite{huang2012differentially, manitara2013privacy, mo2017privacy} failed to achieve average consensus, as confirmed by the numerical simulation results in Fig. \ref{comparison_Zhenqi_Huan_WPES}, Fig. \ref{comparison_Nicolaos_Manitara_ECC}, and Fig. \ref{comparison_Yilin_Mo}, respectively.

\subsection{Hardware Experiments}
\begin{figure}
	\begin{center}
		\includegraphics[width=0.45\textwidth]{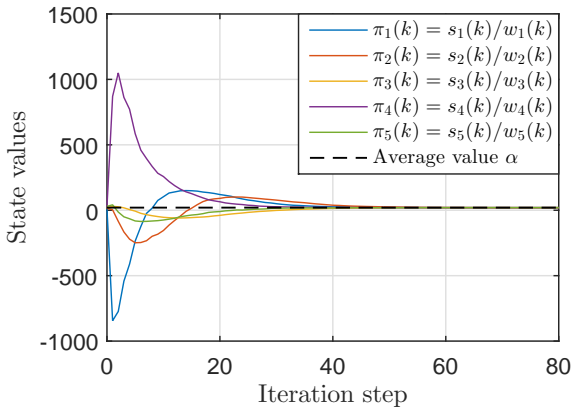}
	\end{center}
	\caption{The evolution of each estimate $\pi_i(k)$ in the hardware implementation on five Raspberry Pi boards.}
	\label{ratio_convergence}
\end{figure}

To confirm the efficiency of our proposed Algorithm 2 in real-world embedded systems, we also implemented our algorithm on five Raspberry Pi boards with 64-bit ARMv8 CPU and 1 GB RAM. In the hardware implementation, the communication was conducted through Wi-Fi based on the ``sys/socket.h'' C library. Paillier encryption/decryption was realized using the ``libpaillier-0.8'' library from \cite{Paillier_Library}. Since in the implementation, it was impossible to start all nodes simultaneously, a counter was introduced on each node which helped nodes to stay on the same pace. For $256$-bit encryption key, the size of the actual packet was $136$ bytes, which included all necessary headers and stuffing bytes. The interaction graph is still Fig. \ref{graph}. The initial values of the five nodes were set to $\{10, 15, 20, 25, 30\}$, which have an average value of $20$. $K$ and $\varepsilon$ were set to $1$ and $0.01$, respectively. The coupling weights $p_{ij}^s(k)$ for $k \leq K$ were chosen from $(-10, \, 10)$. For each encryption, the average processing latency was 12.15 ms, which is acceptable for most real-time control systems. The implementation result is given in Fig. \ref{ratio_convergence}, which shows that for each node $i$, the estimate $\pi_i$ converged to the exact average value $\alpha=20$.

\section{Conclusions}

We proposed a privacy-preserving average consensus approach for directed graphs. Different from existing differential-privacy based approaches which inject noise to exchanged states and thus compromise accuracy, we leverage the inherent robustness of consensus dynamics to embed privacy in random coupling weights, which guarantees consensus to the exact value. Furthermore, the approach can be combined with encryption to protect against eavesdroppers which fail all correlated-noise based privacy-preserving average consensus approaches. Experimental results on embedded micro-controllers confirm that the computational burden is manageable on resource-restricted systems.

\appendix

{\it Proof of Lemma 1}: Given $\mathbf{w}(0)=\mathbf{1}$ and $\mathbf{P}_{w}(k) = \mathbf{I}$ for $k=0,1,\ldots, K$, it is easy to obtain $w_i(k)=1$ for $k \leq K+1$ from (\ref{our_algorithm_I}).

Since $\mathbf{w}(K+1)=\mathbf{1}$ holds, from (\ref{Algorithm_I_second_half}) we have
\begin{equation}\label{Lemma_1_01}     
\begin{aligned}
\mathbf{w}(K+l+1) = \mathbf{\Phi}_w(K+l:K+1) \mathbf{1}
\end{aligned}
\end{equation}	
for $l \geq 1$. Define $\delta(l)$ as
\begin{equation}\label{Lemma_1_02}     
\begin{aligned}
\delta(l) & \triangleq \min_{1 \leq i \leq N}{w}_i(K+l+1) \\
 & = \min_{1 \leq i \leq N} [\mathbf{\Phi}_w(K+l:K+1) \, \mathbf{1}]_i
\end{aligned}
\end{equation}	
for $l \geq 1$. To prove $w_i(k) \geq \varepsilon^N$ for $k \geq K+2$, i.e., $w_i(K+l+1) \geq \varepsilon^N$ for $l \geq 1$, it suffices to prove $\delta(l) \geq \varepsilon^N$ for $l \geq 1$. Next, we divide the proof
into two parts: $1 \leq l\leq N$ and $l \geq N+1$.

\textbf{Part 1:} $\delta(l) \geq \varepsilon^N$ for $1 \leq l\leq N$. It is easy to verify that the following relationship exists
\begin{equation*}    
\begin{aligned}
& [\mathbf{\Phi}_w(K+l:K+1)]_{ii} = [\mathbf{P}_{w}(K+l) \cdots \mathbf{P}_{w}(K+1)]_{ii} \\
\geq & [\mathbf{P}_{w}(K+l)]_{ii} \, [\mathbf{P}_{w}(K+l-1) \cdots \mathbf{P}_{w}(K+1)]_{ii}\\
\geq & \varepsilon \, [\mathbf{\Phi}_w(K+l-1:K+1)]_{ii}
\end{aligned} 
\end{equation*}
Given $[\mathbf{\Phi}_w(K+1:K+1)]_{ii}=[\mathbf{P}_{w}(K+1)]_{ii} \geq  \varepsilon$, we have $[\mathbf{\Phi}_w(K+l:K+1)]_{ii} \geq  \varepsilon^l$. So it follows naturally that
\begin{equation*}
[\mathbf{\Phi}_w(K+l:K+1) \, \mathbf{1}]_i \geq [\mathbf{\Phi}_w(K+l:K+1)]_{ii} \geq \varepsilon^l \geq \varepsilon^N
\end{equation*}
holds for $i=1,\ldots, N$ and $1 \leq l\leq N$, meaning that $\delta(l) \geq \varepsilon^N$ holds for $1 \leq l\leq N$.

\textbf{Part 2:} $\delta(l) \geq \varepsilon^N$ for $l\geq N+1$. Given  $l\geq N+1$ and $\mathbf{\Phi}_w(K+l:K+l-N+1) =\mathbf{P}_{w}(K+l) \cdots \mathbf{P}_{w}(K+l-N+1)$, 
we have the following relationship according to (\ref{Algorithm_I_second_half})
\begin{equation*}
\mathbf{w}(K+l+1) =\mathbf{\Phi}_w(K+l:K+l-N+1) \, \mathbf{w}(K+l-N+1)
\end{equation*}
Letting $\mathbf{w}(K+l-N+1)$ be a standard basis vector $\mathbf{e}_j$ where the $j$-th entry of $\mathbf{e}_j$ is $1$ and all other entries are $0$, we have 
\begin{equation*}
w_i(K+l+1) =[\mathbf{\Phi}_w(K+l:K+l-N+1)]_{ij}
\end{equation*}
Next, we prove 
\begin{equation*}
[\mathbf{\Phi}_w(K+l:K+l-N+1)]_{ij} \geq \varepsilon^N
\end{equation*}
for $1 \leq i, \, j \leq N$, which is equivalent to $w_i(K+l+1) \geq \varepsilon^N$ under $\mathbf{w}(K+l-N+1)=\mathbf{e}_j$.

If $i=j$ holds, given $\mathbf{w}(k+1) =\mathbf{P}_{w}(k) \mathbf{w}(k)$ and $[\mathbf{P}_{w}(k)]_{ii}\geq \varepsilon$ for $k \geq K+1$, we have $w_i(k+1) \geq \varepsilon \, w_i(k)$. Using the fact $w_i(K+l-N+1)=1$, one can easily verify $w_i(K+l+1) \geq \varepsilon^N$, implying $[\mathbf{\Phi}_w(K+l:K+l-N+1)]_{ii} \geq \varepsilon^N$ for $1 \leq i \leq N$.

If $i \neq j$ holds, since the graph $\mathcal{G}$ is strongly connected, there must exist a path from node $j$ to node $i$. Denote the path as $v_1 \shortrightarrow v_2 \shortrightarrow \cdots \shortrightarrow v_r$ where $v_1=j$, $v_r=i$, and $r \leq N$. Since $[\mathbf{P}_{w}(K+l-N+k)]_{v_{k+1} v_{k}}\geq \varepsilon$ holds for $1 \leq k \leq r-1$, we have
\begin{equation*}
w_{v_{k+1}}(K+l-N+k+1) \geq \varepsilon w_{v_{k}}(K+l-N+k)
\end{equation*}
Given $w_{v_{1}}(K+l-N+1)=w_{j}(K+l-N+1)=1$, 
\begin{equation*}
w_{v_{r}}(K+l-N+r)=w_i(K+l-N+r) \geq \varepsilon^{r-1}
\end{equation*}
holds. Further taking into account the facts $[\mathbf{P}_{w}(k)]_{ii} \geq \varepsilon$ for $k \geq K+1$, we have 
\begin{equation*}
w_i(K+l+1) \geq \varepsilon^{N-r+1} w_i(K+l-N+r) \geq \varepsilon^{N-r+1} \varepsilon^{r-1} = \varepsilon^N
\end{equation*}
which implies 
\begin{equation*}
[\mathbf{\Phi}_w(K+l:K+l-N+1)]_{ij} \geq \varepsilon^N
\end{equation*}
for $1 \leq i \neq j \leq N$.

Therefore, we have $[\mathbf{\Phi}_w(K+l:K+l-N+1)]_{ij} \geq \varepsilon^N$ for $1 \leq i, \, j \leq N$. Since $l\geq N+1$ holds and $\mathbf{P}_{w}(k)$ is column stochastic, 
\begin{equation*}
\mathbf{\Phi}_w(K+l-N:K+1) = \mathbf{P}_{w}(K+l-N) \cdots \mathbf{P}_{w}(K+1)
\end{equation*}
is also column stochastic. Further taking into account the fact
\begin{equation*}
\begin{aligned}
& \mathbf{\Phi}_w(K+l:K+1) \\
= \, & \mathbf{\Phi}_w(K+l:K+l-N+1) \mathbf{\Phi}_w(K+l-N:K+1)
\end{aligned}
\end{equation*}
leads to 
\begin{equation*}
[\mathbf{\Phi}_w(K+l:K+1)]_{ij} \geq \varepsilon^N
\end{equation*}
for $1 \leq i, \, j \leq N$. So it can be easily verified that 
\begin{equation*}
[\mathbf{\Phi}_w(K+l:K+1) \, \mathbf{1}]_i \geq N \varepsilon^N \geq \varepsilon^N
\end{equation*}
holds for $i=1,\ldots, N$, meaning that $\delta(l) \geq \varepsilon^N$ holds for $l\geq N+1$.

Therefore, $\delta(l) \geq \varepsilon^N$ holds for $l \geq 1$, which implies $w_i(K+l+1) \geq \varepsilon^N$ for $l \geq 1$. In conclusion, we have $w_i(k)=1$ for $k \leq K+1$ and $w_i(k) \geq \varepsilon^N$ for $k \geq K+2$.   \hfill{$\blacksquare$}

\section*{Acknowledgment}
The authors would like to thank Chunlei Zhang and Muaz Ahmad for valuable discussions and help with experimental verification.

\bibliographystyle{IEEEtran}
\bibliography{abbr_bibli}

\begin{IEEEbiographynophoto}{Huan Gao}
	(S'16) was born in Shandong, China. He received the B.S. and M.Sc. degrees in automation and control theory and control engineering from Northwestern Polytechnical University, Shaanxi, China, in 2011 and 2015, respectively. He is currently working toward the Ph.D. degree in the Department of Electrical and Computer Engineering, Clemson University, Clemson, SC, USA. His current research focuses on dynamics of pulse-coupled oscillators and cooperative control of multi-agent systems.
\end{IEEEbiographynophoto}

\begin{IEEEbiographynophoto}{Yongqiang Wang} (SM'13) was born in Shandong, China. He received the B.S. degree in Electrical Engineering \& Automation, the B.S. degree in Computer Science \& Technology from Xi'an Jiaotong University, Shaanxi, China, in 2004. He received the M.Sc. and the Ph.D. degrees in Control Science \& Engineering from Tsinghua University, Beijing, China, in 2009. 
	
From 2007-2008, he was with the University of Duisburg-Essen, Germany, as a visiting student. He was a Project Scientist at the University of California, Santa Barbara. He is currently an Assistant Professor with the Department of Electrical and Computer Engineering, Clemson University, Clemson, SC, USA. His research interests are cooperative and networked control, synchronization of wireless sensor networks, systems modeling and analysis of biochemical oscillator networks, and model-based fault diagnosis. He received the 2008 Young Author Prize from IFAC Japan Foundation for a paper presented at the 17th IFAC World Congress in Seoul.
\end{IEEEbiographynophoto}

\end{document}